\documentclass[
  journal=largetwo,
  manuscript=article-type,
  year=2020,
  volume=37,
]{cup-journal}
\pdfoutput=1
\usepackage{amsmath}
\usepackage{graphicx,subfigure}

\usepackage{txfonts}
\usepackage{amsmath}	
\usepackage{amssymb}	
\usepackage{indentfirst}
\usepackage{adjustbox}
\usepackage{supertabular,booktabs}
\usepackage{rotating}
\usepackage{nccmath}
\usepackage[section]{placeins}
\usepackage[normalem]{ulem}
\usepackage{xcolor}
 \usepackage{mathrsfs}
\usepackage{booktabs}

\title{A monitoring campaign (2013-2020) of ESA's Mars Express to study interplanetary plasma scintillation}

\author{P. Kummamuru}
\affiliation{School of Natural Sciences, University of Tasmania, Private Bag 37, Hobart, TAS 7005, Australia}
\email[P. Kummamuru]{pradyumna.kummamuru@utas.edu.au}

\author{G. Molera Calv\'es}
\affiliation{School of Natural Sciences, University of Tasmania, Private Bag 37, Hobart, TAS 7005, Australia}

\author{G. Cim\`o}
\affiliation{Joint Institute for VLBI-European Research Infrastructure Consortium, Oude Hogeveendijk 4, 7991, PD Dwingeloo, The Netherlands}

\author{S.V. Pogrebenko}
\affiliation{Joint Institute for VLBI-European Research Infrastructure Consortium, Oude Hogeveendijk 4, 7991, PD Dwingeloo, The Netherlands}

\author{T.M. Bocanegra-Baham\'on}
\affiliation{Jet Propulsion Laboratory, Pasadena, CA, USA}

\author{D.A. Duev}
\affiliation{California Institute of Technology, Pasadena, CA, USA}

\author{M.D. Md Said}
\affiliation{Joint Institute for VLBI-European Research Infrastructure Consortium, Oude Hogeveendijk 4, 7991, PD Dwingeloo, The Netherlands}

\author{J. Edwards}
\affiliation{School of Natural Sciences, University of Tasmania, Private Bag 37, Hobart, TAS 7005, Australia}

\author{M. Ma}
\affiliation{Shanghai Astronomical Observatory, 80 Nandan Road, Shanghai, People’s Republic of China}

\author{J. Quick}
\affiliation{Hartebeesthoek Radio Astronomy Observatory, Krugersdorp, South Africa}

\author{A. Neidhardt}
\affiliation{Technical University of Munich, Research Facility Satellite Geodesy, Geodetic Observatory Wettzell, Sackenrieder Str. 25, D-93444 Bad K\"{o}tzting, Germany}

\author{P. de Vicente}
\affiliation{Observatorio de Yebes (IGN), Yebes, Guadalajara, Spain}

\author{R. Haas}
\affiliation{Chalmers University of Technology, Onsala Space Observatory, G\"{o}teborg, Sweden}

\author{J. Kallunki}
\affiliation{Aalto University Mets\"{o}hovi Radio Observatory, Kylm{\"a}l{\"a}, Finland}

\author{G. Maccaferri}
\affiliation{National Institute for Astrophysics, RadioAstronomy Institute, Radio Observatory Medicina, Italy}

\author{G. Colucci}
\affiliation{E-geos S.p.A, Space Geodesy Center, Italian Space Agency, Matera, Italy}

\author{W. J. Yang}
\affiliation{Xinjiang Astronomical Observatory, Chinese Academy of Sciences, Urumqi, People's Republic of China}

\author{L. F. Hao}
\affiliation{Yunnan Astronomical Observatory, Chinese Academy of Sciences, Kunming, People's Republic of China}

\author{S. Weston}
\affiliation{Institute for Radio Astronomy \& Space Research (IRASR)
School of Engineering, Computer and Mathematical Sciences, Faculty of Design and Creative Technologies, Auckland University of Technology, New Zealand
}

\author{M. A. Kharinov}
\affiliation{Institute of Applied Astronomy of Russian Academy of Sciences, St Petersburg, Russia}

\author{A. G. Mikhailov}
\affiliation{Institute of Applied Astronomy of Russian Academy of Sciences, St Petersburg, Russia}

\author{T. Jung}
\affiliation{Korea Astronomy \& Space Science Institute, 776 Daedeok-daero, Yuseong-gu, Daejeon, South Korea}


\begin{document}

\begin{abstract}
The radio signal transmitted by the Mars Express (MEX) spacecraft was observed regularly between the years 2013-2020 at X-band (8.42 GHz) using the European Very Long Baseline Interferometry (EVN) network and University of Tasmania's telescopes. We present a method to describe the solar wind parameters by quantifying the effects of plasma on our radio signal. In doing so, we identify all the uncompensated effects on the radio signal and see which coronal processes drive them. From a technical standpoint, quantifying the effect of the plasma on the radio signal helps phase referencing for precision spacecraft tracking. The phase fluctuation of the signal was determined for Mars' orbit for solar elongation angles from 0 - 180 deg. The calculated phase residuals allow determination of the phase power spectrum. The total electron content (TEC) of the solar plasma along the line of sight is calculated by removing effects from mechanical and ionospheric noises. The spectral index was determined as $-2.43 \pm 0.11$ which is in agreement with Kolomogorov's turbulence. The theoretical models are consistent with observations at lower solar elongations however at higher solar elongation ($>$160 deg) we see the observed values to be higher. This can be caused when the uplink and downlink signals are positively correlated as a result of passing through identical plasma sheets. 
\end{abstract}


\section{Introduction}
The last several decades have seen a significant number of spacecraft launched to explore the Solar System. Techniques like Very Long Baseline Interferometry (VLBI) and Doppler spacecraft tracking have progressively been used over the same period for different space science missions. The Planetary Radio Interferometry and Doppler Experiment (PRIDE) program was developed by the Joint Institute for Very Long Baseline Interferometry European Research Infrastructure Consortium (JIVE) which uses the VLBI and Doppler techniques to conduct radio science experiments for scientific and orbit determination purposes. \cite{calves2021high} describes the software which was key to single dish data processing of spacecraft signals with VLBI telescopes. The software was key in the observations of Venus Express' (VEX)~\citep{duev2012spacecraft} and MEX Phobos flyby~\citep{duev2016planetary} for ultra-precise orbit determination. PRIDE has been used in several other science experiments; the study of interplanetary phase scintillation using spacecraft signals from VEX~\citep{calves2014observations}, noise budget estimation of the MEX Phobos flyby \citep{bocanegra2017planetary}, radio occultation experiment with the ESA's Venus express (VEX) to study Venus' atmosphere~\citep{bocanegra2019venus}. The technique will play a crucial role in the upcoming European Space Agency's 
(ESA) Jupiter Icy Moons Explorer (JUICE) mission scheduled to launch in 2023.

The characterization of interplanetary plasma is a crucial component for achieving high-precision astrometry with the PRIDE technique. The presence of interplanetary plasma is a result of the outflow of ionized particles from the Sun known as solar winds. Solar winds are broadly classified based on their speeds into slow and fast solar winds. Slow solar winds are characterized by speeds between 300-500 km/s and higher density of $~ 10^{7} cm^{-3}$. Their origins are not fully understood but some of the major models explaining the origin of the slow solar winds are the coronal flux tube expansion phenomenon (Eg.~\cite{pinto2017multiple}) and the interchange magnetic reconnection (Eg.~\cite{edmondson2012role}). Fast solar winds have speeds ranging from 600-800 km/s with lower densities compared to slow solar winds with a value around $5 \times 10^{5} cm^{-3}$. The fast solar winds originate from the coronal holes, regions of open magnetic field lines that propel matter into space~\citep{hassler1999solar}.

A spacecraft communications telemetry signal is affected by multiple factors along the propagation path including the motion of the spacecraft, the characteristic's of the antenna, the helioplasma, Earth's ionosphere, and spacecraft and antenna system noises. These affect the observables of the spacecraft signal including signal frequency, phase and amplitude. The amplitude of a signal is significantly disrupted when the line of sight is close to the Sun \citep{manoharan1995solar} such as a solar conjunction. The automatic gain control (AGC) is always switched on in our experiments. The AGC provides a controlled received amplitude facilitating easier signal processing with less changes in the dynamic range. Due to the two aforementioned reasons, the amplitude of the signal is not a suitable metric in comparison to the more precisely detected phase. The solar wind introduces frequency \citep{wexler2019spacecraft} and phase fluctuations in the signal. These fluctuations become larger when observations are closer to the Sun or during coronal mass ejections events. We observe spacecraft downlink signals operating in a coherent communication mode where the spacecraft generates a downlink signal coherent with the transmitted ground station uplink, offering increased phase stability. We then analyze the phase fluctuations of the spacecraft carrier signal to characterize the impact of interplanetary plasma. 

\cite{calves2014observations} determined the phase fluctuation indices of VEX's telemetry signal along Venus' orbit between the years of 2009-2013 for solar elongation angles over the range of 0-45 deg. In this study, we use the telemetry signal of MEX to observe the phase fluctuations over a larger extent of solar elongation angles 0-180 deg to get a more extensive coverage. Using the phase fluctuation spectrum we determined from our Doppler observations, we calculate the TEC along the sightlines to Mars. The extended observing campaign of nearly 3 orbital periods of Mars allows us to confirm the solar density profile across the entire span of solar elongation. Observations at higher solar elongations are essential to extend and improve current theoretical models of the total electron content of interplanetary plasma. 

The relevance of studying the phase spectrum of plasma extends to the PRIDE's experiment of tracking spacecraft using the VLBI-phase referencing technique. In this technique, telescopes, while tracking, switch sources between the spacecraft and flux calibrators which are ideally separated by only a few degrees. When alternating from one target to the other, one of the key parameters is the nodding cycle, the time spent on observing the spacecraft and the calibrator and the switching time between targets. The nodding interval between sources has to be adjusted so that the path length change due to the phase errors is \(< \lambda/4\) where \(\lambda\) is the wavelength of the signal \citep{beasley1995vlbi}. The measured phase fluctuations are a symptom of noise errors introduced in the radio signal due to the propagation media; quantifying this would allow us to select optimum nodding cycles. This consequentially enables the precise determination of the spacecraft's state vectors. 

In the next part of the paper, we discuss the theory of phase scintillation spawning through the interplanetary plasma region. This is followed up by an overview of the observations taken over the full campaign and the methodologies used for the data processing. In section four, we demonstrate our results and compare them with previous results. In section five, we discuss the implications of our results and recognize the avenues for improvement for future studies.

\section{Theory}

The upper atmosphere of the Sun, the corona, is responsible for releasing a stream of plasma which is the solar wind. The solar wind plasma overlayed with the heliospheric magnetic field \citep{owens2013heliospheric} permeates through the interplanetary medium. When radio signals are sent to and from spacecraft they pass through the interplanetary plasma and as a consequence we observe fluctuations in the signal. The scattering regime of the plasma is modelled in different ways based on geometry and distance and can be understood to have weak and strong scattering zones. A majority of the solar wind scattering is in the weak regime while it enters into the strong regime closer to the Sun \citep{narayan1992physics}. In the weak scattering domain, the  fluctuations are caused due to the electron density variations in the solar wind which scatter the radio waves. The distortions in the phase fronts are dependent on the size of the plasma irregularities which can be both diffractive and refractive with the latter associated with smaller irregularities \citep{boyde2022lensing}. \cite{conroy2022statistical} delves further on distinguishing refractive and diffractive phase scintillation. 

Consolidating a quantitative relationship between the scintillation and the plasma density will help us get a better insight into the solar wind structure and thereby, the corona. The fluctuations we observe in our radio signal is a consequence of large-scale structure of the solar wind \citep{schwenn1990large}.  We look at how the expanse of the inhomogeneous medium translates to the electron density across the line of sight of our observations.  

The phase scintillation provides information on the full range of scale sizes for electron density variations at different distances and is typified by the refractive index. The inhomogenous medium of plasma spans across thousands of kilometres \citep{yakovlev2002space} and is characterized by a spectrum of refractive coefficients given by

\begin{ceqn}
\begin{align}
 \phi_{n}(\kappa) = 0.033c_{n}^{2}(\kappa^{2} + \kappa_{0}^{2})^{-\alpha/2}\exp\left({-\frac{\kappa^{2}}{\kappa_{m}^{2}}}\right)   
\end{align}
\end{ceqn}
 
where \(k=2\pi/\lambda\) represents the spatial wave number, \(\kappa\) is the scale of the plasma irregularity that the signal traverses, \(\kappa_{m}\) and \(\kappa_{0}\) are spatial wave numbers in the outer and inner scales of the refractive index regularities, \(\alpha\) is the plasma irregularity's spatial spectrum index, and \(c_{n}\) is the plasma irregularity's structural coefficient. 
The radio wave fluctuations are proportional to the spatial spectrum of the refractive index with the fluctuations in the phase being more pronounced than the amplitude. The variance of the phase of the fluctuations is calculated as 

\begin{ceqn}  
\begin{align}
 \sigma_{\psi}^2 = (2{\pi}\nu)^{2}\int_{0}^{L}\int_{0}^{\kappa_{0}} \phi_{n}(\kappa){\kappa} d{\kappa}dx
\end{align}
\label{eqn:factor}
\end{ceqn}
  
where $\nu$ is the wavenumber of the wave travelling through vacuum. We integrate over the satellite to Earth path length (L) and the extent of the wavenumber of the plasma inhomogeneity spectrum (\(\kappa\)) from 0 to \(2{\pi}\Lambda_{0}^{-1}\) (\(\Lambda=10^{6} km\) is the outer turbulence scale) \citep{yakovlev2002space}.

In this work, we determine the phase scintillation as the standard deviation of the phase residual values. The phase residuals are extracted from fine Doppler detection of the carrier signal and is the quantity obtained after compensating residual phase rotation. The phase residuals are used to construct the phase power spectrum which gives us a qualitative and quantitative insight into the interplanetary plasma.  

Another measure of the effects of the propagation on the radio signal is the total electron content (TEC), expressed in electrons per square metre; one TEC unit (tecu) is \(10^{16} electrons/m^{2}\). The TEC is calculated as a line integral along the line of sight from the Earth to a spacecraft as follows

\begin{ceqn}  
\begin{align}
  TEC=\frac{1}{tecu}\sum_{s=E}^{SC}N_{e}(s)\cdot{d_{seg}}
\end{align}
\end{ceqn}
 
where E represents the position of the Earth in a 2D map and SC the position of the spacecraft, $N_e$ is the electron column density function of the solar wind with respect to the Sun normalized by $tecu$, and $d_{seg}$ is the distance increment determined by the number of intervals of the line of sight between observer-target. The electron density is integrated along the path length and is given for the nominal, slow, and fast solar winds as follows~\citep{ando2015internal}:

\begin{ceqn}
\begin{align}
     {N_{nominal}=5\times10^{6}\cdot(AU/d)^{2} m^{-3},}
\end{align}
\end{ceqn}

 \begin{ceqn}
\begin{multline}
     N_{slow}=(4.1\cdot{s_{dist}}^{-2}+23.53\cdot{s_{dist}}^{-2.7})\cdot10^{11} + \\ (1.5\cdot{s_{dist}}^{-6}+2.99\cdot{s_{dist}}^{-16})\cdot10^{14},
\end{multline}
\end{ceqn}

\begin{ceqn}
\begin{multline}
     N_{fast}=(1.155\cdot{s_{dist}}^{-2}+32.2\cdot{s_{dist}}^{-4.39}+ \\ 3254\cdot{s_{dist}}^{-16.25})\cdot10^{11},
\end{multline}
\end{ceqn}

where d is the solar offset (in metres) and $s_{dist}$ is the ratio of solar offset and the radius of the Sun.

The signal is also affected by ionospheric and tropospheric contributions along its path \citep{crane1977ionospheric,karasawa1988new}. The ionospheric induced phase delay is given by 

\begin{ceqn}
\begin{align}
      I=-\frac{f_{p} \times TEC}{f^{2}}
\end{align}
\end{ceqn}

where $f_{p}=8.98\sqrt{N}$ (Hz) is the frequency of the plasma medium \citep{davies1990ionospheric} and $f$ is the transmission frequency. The phase delay of the signal in cycles is determined by dividing the ionospheric path contribution by the wavelength, \(L=I/\lambda\).

Radio links are subject to tropospheric scintillation due to refractive index fluctuations. The refraction has both wet and dry components with the latter largely dominant (90\%) and well correlated with the atmospheric pressure making it easy to determine \citep{jin2007seasonal}. The delay due to the wet component is highly variable because of the rapidly fluctuating water vapour content in the atmosphere. The accurate modelling of the wet component contribution relies on the use of high quality radiosondes like the Topex \citep{keihm1995topex} which is infeasible for every area of science unless it is high-precision GPS work, thus we rely on mathematical models. However, the signal attenuation increases at higher microwave frequencies ($>$10 GHz) and as a consequence we lose any benefits of atmospheric refraction at this stage \citep{vasseur1999prediction,1697229}. The atmospheric delay is higher at lower elevations because of how it is proportional to $1/\sin{\epsilon}$ where $\epsilon$ is the elevation angle \citep{macmillan1994evaluation}. However, the root mean square ($rms$) of tropospheric induced phase fluctuation is negligible compared to the solar wind value \citep{acosta2010path,holdaway1995data} while we tend to ignore the lower elevation observations ($<$10 deg) due to tropospheric saturation.

\section{Observational Summary}
\subsection{Observations}
The observing campaign was held across 303 epochs with a total of 504 sessions using 22 different radio telescopes around the world. Table \ref{tab:sefd} shows a description of antennas, station code, geographical location, the System Equivalent Flux Density (SEFD), and the dish size used in the observations. These antennas all are equipped with a receiver capable to operate at X-band frequencies. Although, the system noise of the antenna (expressed as the SEFD) varies significantly among them. The variety of antennas does not impact the data output obtained with the quality of our measurements as seen later in Figure~\ref{fig:carrsnr}. 

\begin{center}
 \begin{table}[ht]
 \begin{tabular}{||c c c c||} 
 \hline
 Antenna (code) & Country & SEFD(Jy) & $\phi$(m)\\ [0.5ex]
 \hline\hline
 Ceduna (Cd) & Australia & 600 & 30 \\ 
 \hline
 Hobart (Ho) & Australia & 2500 & 26 \\
 \hline
 Katherine (Ke) & Australia & 3500 & 12 \\
 \hline
 Yarragadee (Yg) & Australia & 3500 & 12 \\
 \hline
 Hobart (Hb) & Australia & 3500 & 12 \\ 
 \hline
 Svetloe (Sv) & Russia & 350 & 32 \\
 \hline
 Zelenchuk (Zc) & Russia & 350 & 32 \\
 \hline
 Badary (Bd) & Russia & 350 & 32 \\
 \hline
 Tianma (T6) & China & 200 & 65 \\
 \hline
 Yebes (Ys) & Spain & 200 & 40 \\
 \hline
 Hartebeesthoek (Ht) & South Africa & 3000 & 15 \\
 \hline
 Warkworth (Ww) & New Zealand & 3500 & 12 \\
 \hline
 Kunming (Km) & China & 1500 & 40 \\
 \hline
 Sheshan (Sh) & China & 1500 & 12 \\
 \hline
 Metsahovi (Mh) & Finland & 3200 & 14 \\
 \hline
 Hartebeesthoek (Hh) & South Africa & 3000 & 26 \\
 \hline
 Onsala (On) & Sweden & 1500 & 20 \\
 \hline
 Wettzell (Wz) & Germany & 750 & 20 \\
 \hline
 Wettzell (Wn) & Germany & 1400 & 13.2 \\
 \hline
 Warkworth (Wa) & New Zealand & 900 & 30 \\
 \hline
 KVN Ulsan (Ku) & South Korea & 1080 & 21 \\[1ex] 
 \hline
\end{tabular}
 \caption{Overview of the telescopes used for our observations of ESA MEX spacecraft, their locations, SEFD values, and the diameter of the main parabolic dish.}
   \label{tab:sefd}
 \end{table}
\end{center}

\begin{figure}[htb]
\centering
\includegraphics[width=\columnwidth,keepaspectratio]{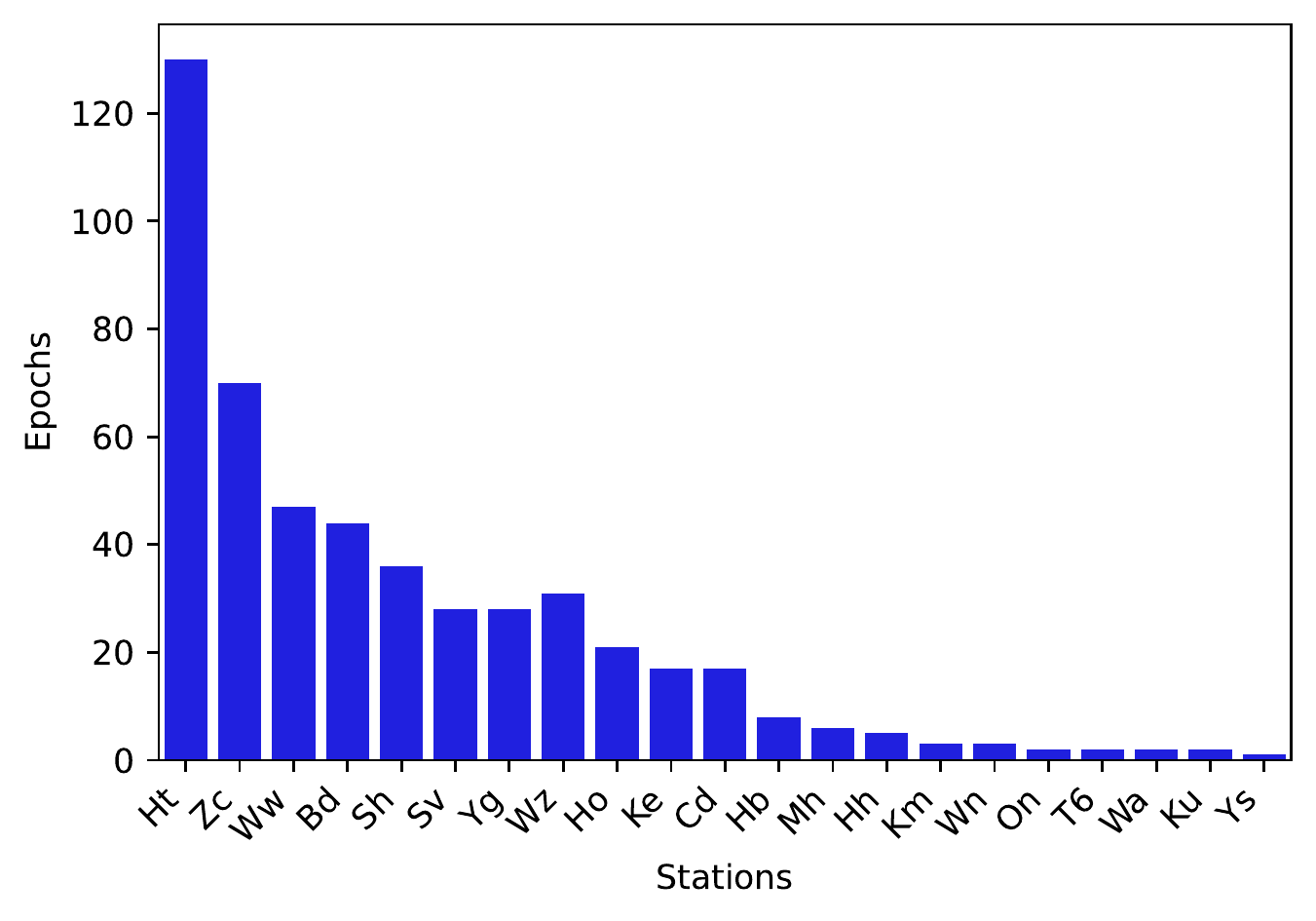}
\caption{The distribution of number of observations conducted with each radio telescope between 2013 and 2020.}
\label{fig:distri}
\end{figure}

The number of observations conducted by each antenna varied depending on the Mars visibility, antenna availability, and transmission times of MEX. The distribution of the amount of observations over the past ten years is given in Figure~\ref{fig:distri}. Most of the data collected in this study were observed with the KVAZAR network of VLBI antennas in Russia, including Svetloe, Zelenschunkya and Badary, Hartebeesthoek in South Africa, the 12m and 30m telescopes from Warkworth, New Zealand \citep{woodburn2015conversion}, and the five radio telescopes operated by the University of Tasmania. All the sessions conducted from 2019 and onwards were conducted with antennas in Hobart, Katherine, Yarragadee, and Ceduna.

\begin{figure}
\centering
\includegraphics[width=\columnwidth,keepaspectratio]{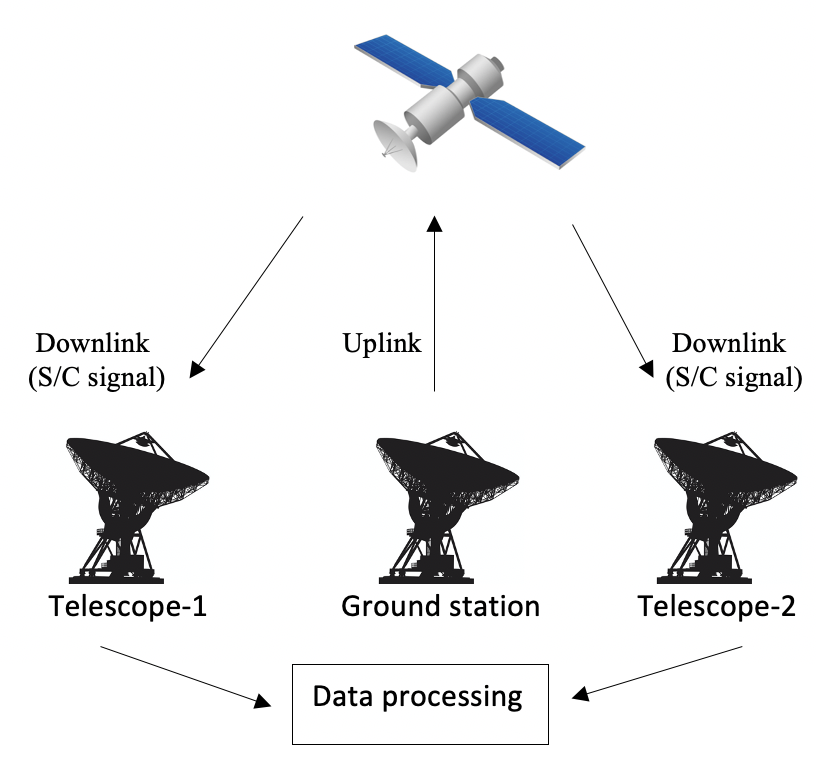}
\caption{Observations conducted with PRIDE use the three way mode, in which the spacecraft is operating in two-way mode with ESTRACK stations and VLBI radio telescopes detect the signal in a third location.}
\label{fig:3way}
\end{figure}

The objective of these sessions was to quantify the phase fluctuations of the radio signal at different solar elongations. The observations covered the period between 2013 to 2020 an equivalent to three orbital periods of Mars. The onboard receiver system of MEX is capable of receiving and transmit radio signals in S-band and X-band utilizing a High Gain Antenna. The communication is either a one-way link using the spacecraft's in-built ultra stable oscillator or a two-way link wherein an initial signal is transmitted from an Earth station which gets locked in the spacecraft and then re-transmitted to Earth \citep{asmar2005spacecraft}. For our observations, we used a three-way mode (Figure \ref{fig:3way}), a variation of the two-way, wherein the transmitting ground station is different to the receiving ground station. The receiving stations are generally equipped with hydrogen masers which provide better phase stability as accurate reference clocks.

Each session was segmented into scans of 19 minutes. This is adequate to deconstruct the phase fluctuations down to a milliradians resolution and keep consistent with \cite{calves2014observations}. We record the broadband radio signal in VLBI specific data format \citep{whitney2009vlbi} in multiple frequency channels with bandwidths of 8, 16 or 32\,MHz depending on the station's back-end configuration. For example, the latest digital back-end at the UTAS stations of Katherine and Hobart specifically supports 32\,MHz bandwidth per channel, recording only linear polarised radio signals. The recorded raw files for each session consist of ten to hundreds of gigabytes depending on the session's length. 

\subsection{Data processing}

The data obtained at individual telescope sites was either sent in its raw state or partially processed using the software pipeline and then sent across to the investigation centre by electronic or physical means. The received raw spacecraft files are processed through a high spectral resolution multi-tone Spacecraft Doppler tracking software (SDtracker)~\citep{calves2021high} to extract the topocentric frequency and residual phase of the spacecraft. The SDtracker comprises three major steps of Software spectrometer (SWspec), the Spacecraft tracker (SCtracker), and digital Phase Locked Loop (dPLL)~\footnote{https://gitlab.com/gofrito/sctracker/} outlined below.

\subsubsection{SWspec}
\begin{figure}

    \centering
    \subfigure[Full-stacked spectra]{\label{fig:fdet}\includegraphics[width=\columnwidth,keepaspectratio]{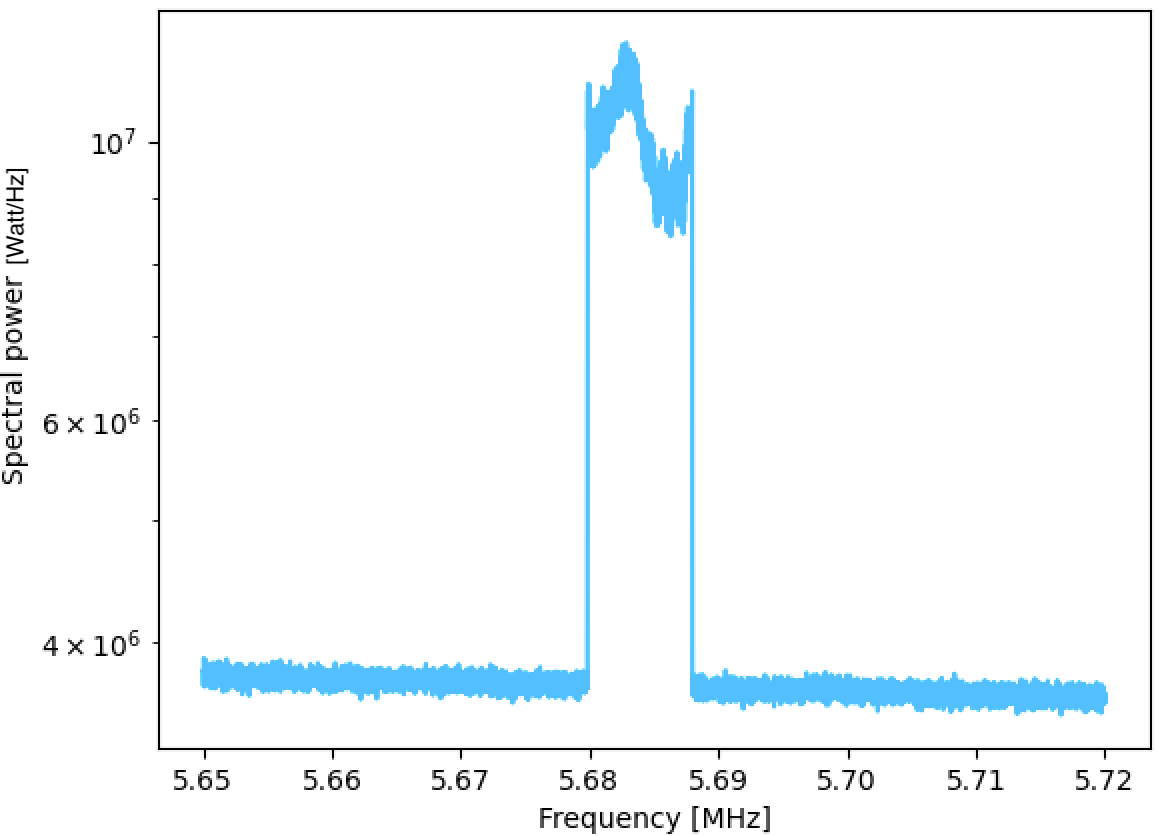}}
  \subfigure[Frequency detections]{\label{fig:fdetect}\includegraphics[width=\columnwidth,keepaspectratio]{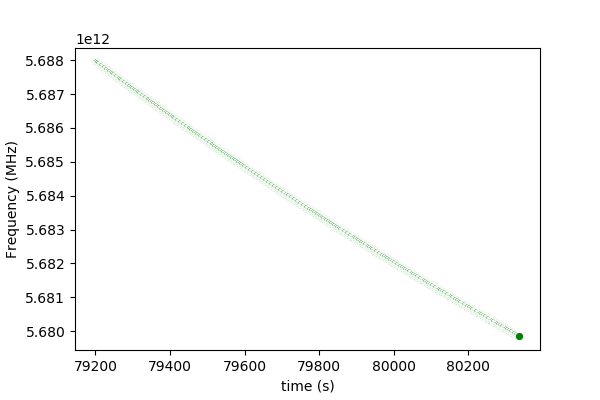}}
   \caption{(a) Detection of the spacecraft carrier signal on the spectrum for a session held at Yarragadee. (b) The Doppler shift of the detected spacecraft tone over the course of the 19 minute scan. We use a sixth order polynomial to fit the shift in the frequency tone.)}
  
   \label{fig:dets}
\end{figure}

The first step in processing the data is to identify the channel containing the spacecraft carrier signal which is easily determined from the pre-existing transmission frequency information. In this step, a time-integration of the scan is performed wherein multiple spectra are generated which is given by \(N=LS/i\) where \(LS\) is the length of scan in seconds and \(i\) is the integration time in seconds. It is important to note here that the integration time used for processing the MEX signal (2\,s) is shorter than that used for the VEX (5s) because the orbit of the MEX is such that it is rotating more rapidly compared to VEX thereby requiring more iterations (spectra) for better resolution. A procedure of window-overlapped add discrete Fourier transform (WOLA-DFT) is then performed on the data to obtain the shift in frequency across the spectrum. The shift is determined using a polynomial fit. Typically, in spacecraft detections we use a sixth-order polynomial fit but \cite{ma2021vlbi} determined that fourth and fifth order profile fits give nearly similar results. Figure \ref{fig:dets} shows an example of the detection of the MEX spacecraft signal and the fitted frequency shift profile.

\subsubsection{Multi-tone tracking and phase-locked loop}

\begin{figure}[hbtp]
    \centering
    \includegraphics[width=\columnwidth,keepaspectratio]{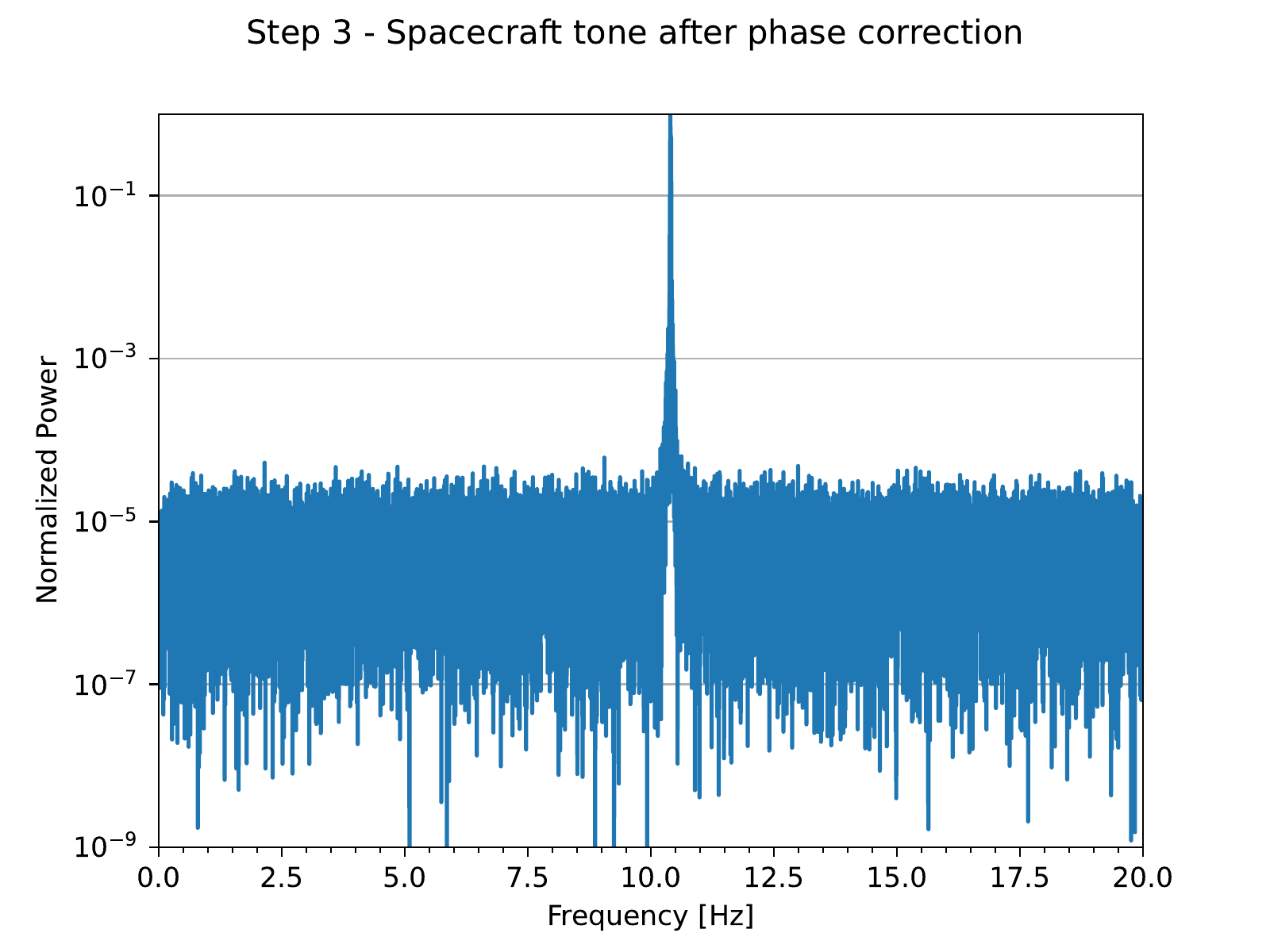}
   \caption{The narrow tone of the MEX carrier signal obtained after the digital phase locked loop.} 
   \label{fig:nato}
\end{figure}

In this step, the phase of the spacecraft carrier tone is stopped and we get the Doppler corrections to resolve the tone to within a narrow mHz level. This is done using a time integration algorithm. From the obtained spectrum, a narrow window is selected around the spacecraft tone and subject to a 2nd order DFT-based algorithm which gives a Doppler-corrected spacecraft tone in a 1-2 kHz narrow bandwidth (Figure \ref{fig:nato}).

The obtained tone is passed to the digital phase locked loop software which runs high precision iterations. The software calculates a new time-integrated spectra at every step, estimates a new set of phase polynomial fit, and then does the phase stopping of the spectra. The output at every step is a new filtered and down-converted signal associated with a residual frequency and phase. The output bandwidth phase of the detections post the dPLL processing has a Doppler noise less than a hundred mHz.

\subsubsection{Scintillation analysis}

In this final step of the data processing, we derive the temporal variation of the phase fluctuations over each scan and use the combined phase residuals of all session to estimate the phase scintillation indices. We start by plotting the residual phases for all the recorded scans. We inspect them looking for outlier, which mostly could be caused by a phase jump on the data. Depending on the spacecraft motion or low SNR the dPLL stage is not capable to unwrap perfectly the phase of the spacecraft. These jumps can sometimes be recovered by reducing the number of FFT points per segment in the dPLL. This may however not work when there is a short gap in the data caused by the recorder. In these outlier cases, we discard the scan and continue with the remaining data set.

The first step is to evaluate if the root mean square (\textit{rms}) of phase fluctuations in radians is consistent in all the scans. The \textit{rms} is actually the total energy of the phase fluctuations and it can be calculated as seen below.

\begin{ceqn}  
\begin{align}
E=\sum_{j}[\sigma^{2}dt]
\end{align}
\end{ceqn}

We can also express the power in time domain terms as:

\begin{ceqn}  
\begin{align}
P=\frac{E}{Ndt}
\end{align}
\end{ceqn}

where $E$ is the total energy, $\sigma$ us the phase residual, $j$ is the time scale spanning across the '$N$' data points and $dt$ is the sampling interval.

From this point, we are converting our time domain samples into frequency domain by using a windowed Fast Fourier Transformation. The phase power spectrum gives us an insight into the large scale structure of the solar wind; showing in which range of frequencies the effect of plasma scintillation (scintillation band) is more prominent compared to the noise band. Determining the phase power spectral density involves a few steps.

We use two different forms of density spectra: windowed and unwindowed. We use the windowed spectra for estimation of the slope and relate it to a typical Kolmogorov spectrum. We use the unwindowed spectra for filtering and estimation of the scintillation \textit{rms} and noise system \textit{rms}.

\begin{ceqn}  
\begin{align}
psp = \mathcal{F}(\sigma)\\
pspw = \mathcal{F}(\sigma \cdot win)
\end{align}
\end{ceqn}

where $pspw$ is the windowed spectra and $psp$ is the unwindowed spectra. The units are expressed in square radians per Hz. We calculate the power spectra for each of the scans and then we stack them all together.

\begin{ceqn}  
\begin{align}
  pspw = \frac{1}{Ns} \cdot \left( \sum_{j} \frac{2\cdot pspw}{BW} \right)
\end{align}
\end{ceqn}

The scintillation caused by the plasma can be determined by doing a first order approximation of the power spectral density given as 

\begin{ceqn}  
\begin{align}
L_{ps}=c+mL_{f}
\end{align}
\end{ceqn}

where \(L_{ps}\) is the average-windowed power spectral, $m$ is the slope of the fitted line, \(L_{f}\) the frequency on logarithmic scale, and $c$ is a constant. The limits of the best fit line are taken where the slope looks linear in the log-log scale. The red line in Figure 6 represents this line of fit. The slope is indicative of the spectral index which represents how the solar wind varies with phase.

We set frequency limits within the spectral density to distinguish between the contributions of scintillation and noise to phase fluctuations. The lower limit frequency for the scintillation band is taken as 0.003 Hz which represents the effective integration time while the upper limit is taken between 0.1 - 0.5 Hz depending on where the system noise band starts to dominate. These limits can be seen in Figure 6. The standard deviation of the phase fluctuation in the noise band determines the system noise level while the standard deviation of the phase fluctuation in the scintillation band (interplanetary plasma) gives the scintillation index.

\section{Results}

\subsection{Analysis of the phase fluctuations}

\begin{figure}[hbt]
    \centering
    \includegraphics[width=\columnwidth,keepaspectratio]{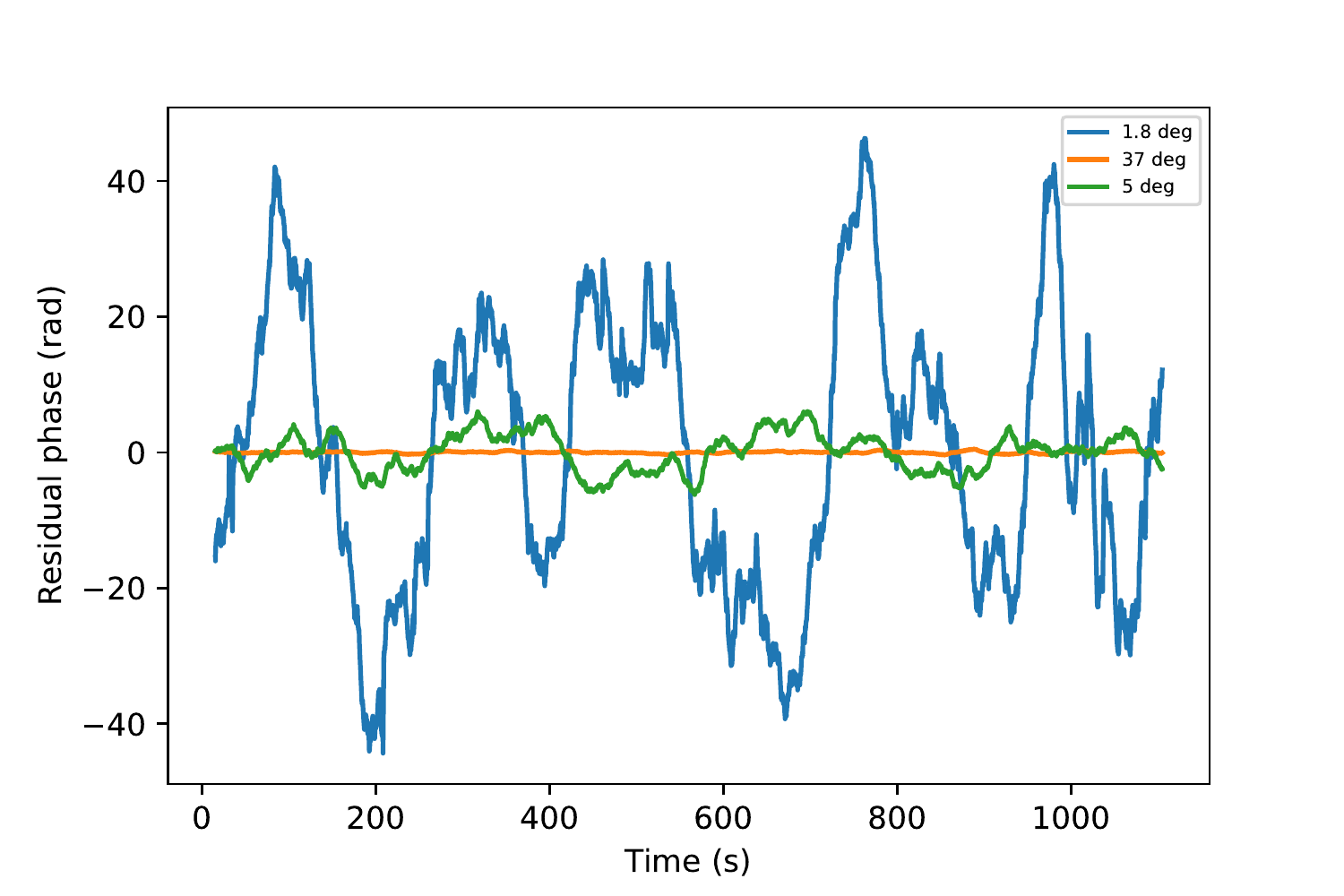}
   \caption{Phase fluctuation at different solar elongations (1.8 deg, 5 deg and 37 deg) at Hartebeesthoek (Ht) on three different epochs (8th June 2015, 9th July 2015 and 8th February 2016 respectively). } 
   \label{fig:pdeg}
\end{figure}

The phase of the spacecraft carrier signal is affected by the interplanetary plasma during both up and down-link transmissions. The scan duration (19m) and number density of electrons are key factors that influence the value of the phase scintillation. Increasing the scan length increases the average phase residual values and for an optimum scan length to capture the long-scale structure of the solar wind we picked 19 minutes; it's the same duration \cite{calves2014observations} chose for the observations of VEX thus allowing for a consistent comparison. The density of electrons increases at lower solar elongations and thus the expected trend of higher phase fluctuations at low solar elongations is consistent with our studies as seen in Figure \ref{fig:pdeg}. The blue line in the figure is an observation at the Hartebeesthoek station from the 8th of August 2016 at a very low solar elongation angle (1.8 deg) which shows a high phase fluctuation while in contrast the green and yellow lines correspond to higher solar elongation angles of 5 deg and 37 deg respectively which demonstrate lesser phase fluctuation. 

\begin{figure}
 \centering
 \includegraphics[width=\columnwidth,keepaspectratio]{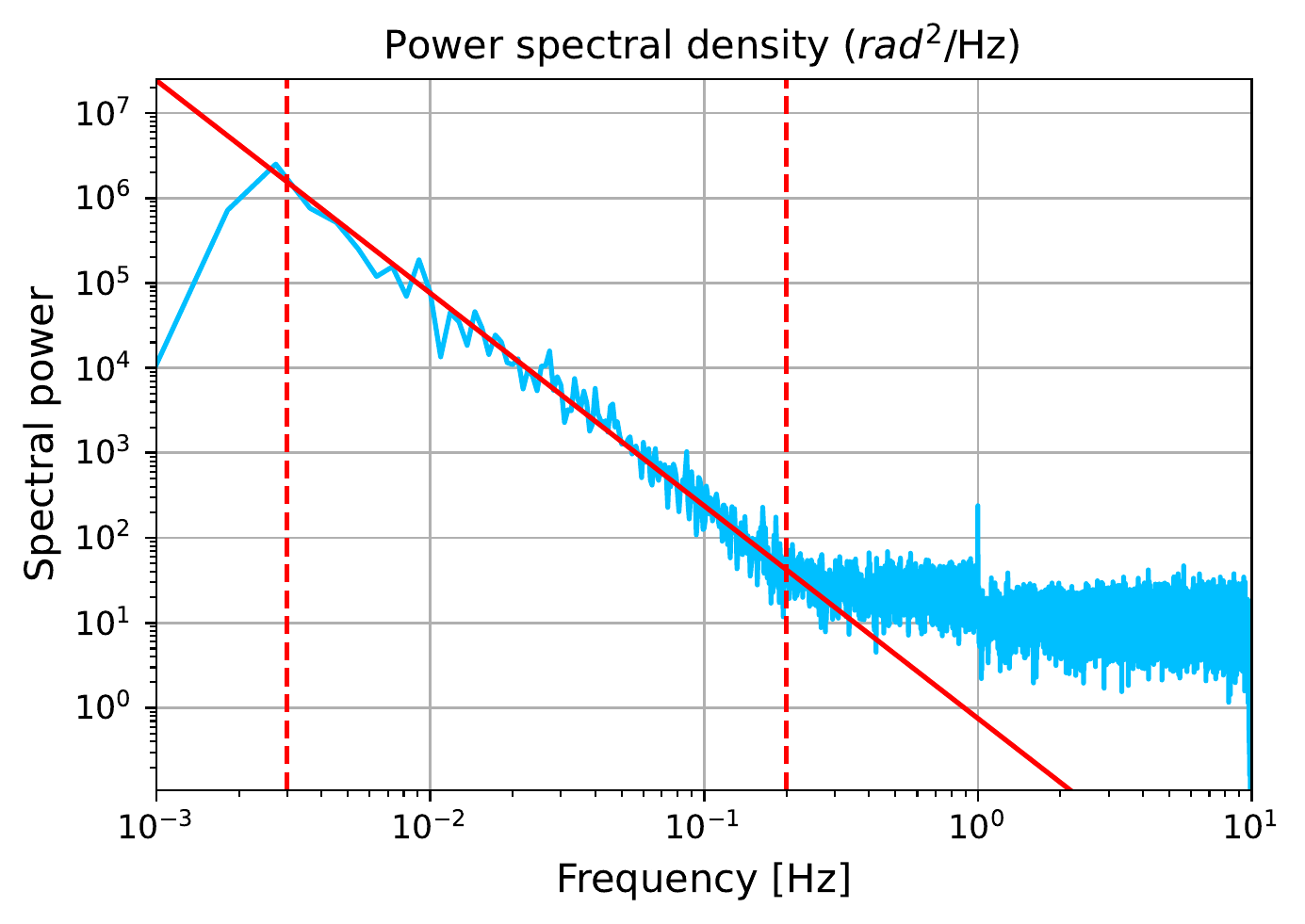}
 \caption{The power spectral density of a session held at Yarragadee on the 13th of April, 2020. The two horizontal dotted red lines encapsulate the scintillation band with the slope of the spectrum's fit (red line) corresponding to -2.431. The region beyond the second (right) dotted line corresponds to the noise band.}
 \label{fig:spd}
\end{figure}

The contributions of the scintillation and the noise are differentiated by plotting a first-order approximation fit of the power spectral density on a logarithmic scale. The slope of the fit is indicative of the spectral index which is consistent with the Kolmogorov power spectrum of turbulence. The values of the slope of the spectral index for our observations ranged from $-$3.208 to -1.374 with a mean value of $-2.43 \pm 0.11$. This is similar to the value of $-2.42 \pm 0.25$ obtained from previous VEX observations \citep{calves2014observations} and the value of $-$2.45 found by \cite{woo1979spacecraft}. The slope is given as $m=1-p$ where $p$ is the Kolmogorov index. The average value of $m = -2.42$ we obtain agrees with the $p=11/3$ from the earlier study of \cite{kolmogorov1991local}. The Figure \ref{fig:spd} shows the phase power spectral density from a session held at Yarragadee. The scintillation band shows us the range of frequencies where the plasma is dominant (0.008 - 0.3 Hz) and the noise band is where the system noise effects start dominating.

\begin{figure}[hbt]
 \centering
 \includegraphics[width=\columnwidth,keepaspectratio]{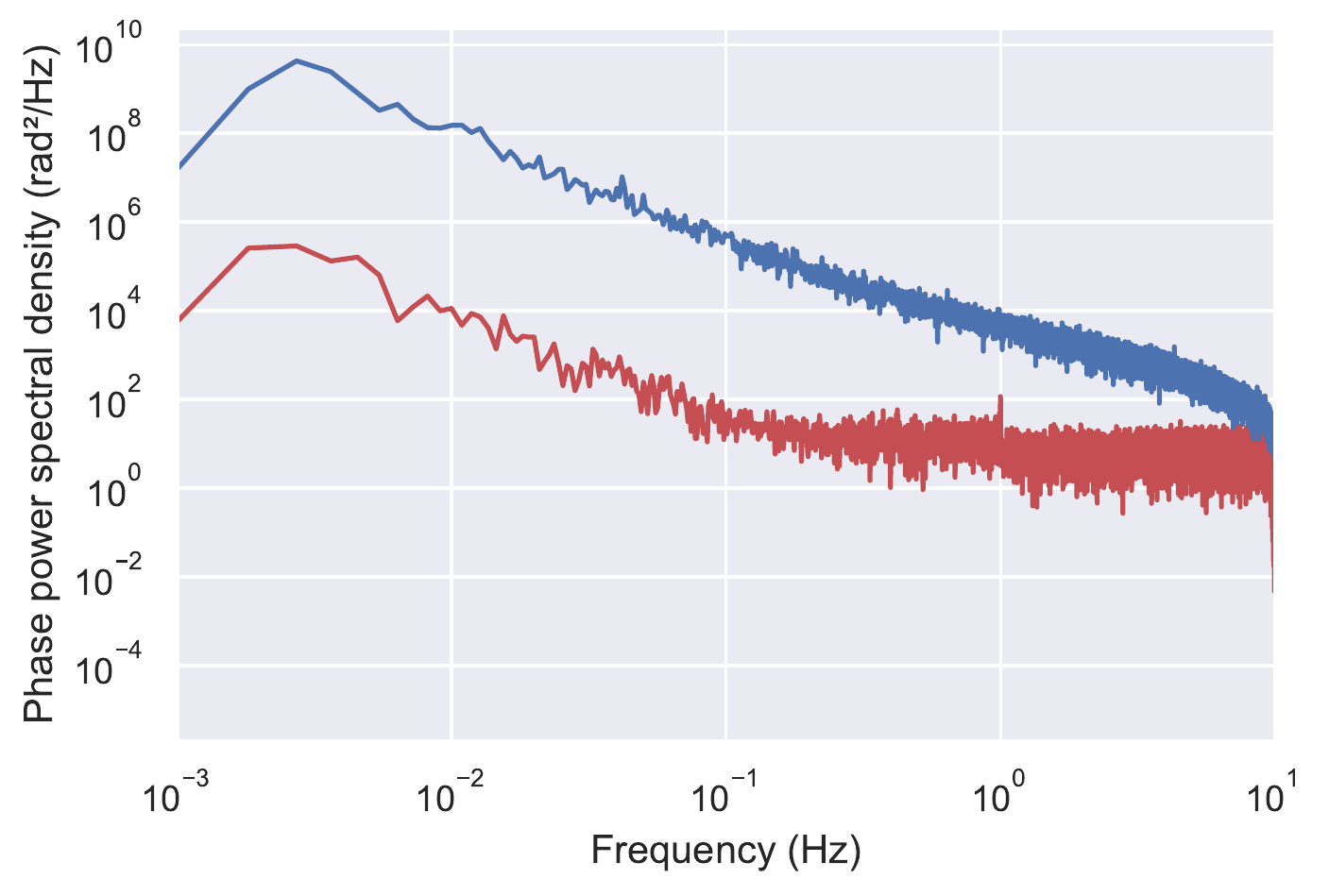}
 \caption{The power spectral density of two sessions held at Yarragadee where the blue spectrum is when the solar elongation was 4.8 deg and the red spectrum is when the solar elongation was 87.3 deg.}
 \label{fig:psdcom}
\end{figure}

The spectral index value calculated for each of the sessions is found to be independent of the scintillation. In Figure \ref{fig:psdcom}, we can distinguish the phase power levels for the low and high solar elongation spectra for the same station Yarragadee. The other noticeable feature is the dominance of the scintillation band over a longer frequency range in our lower solar elongation observation.

\subsection{Scintillation analysis of 3 orbital periods}

\begin{figure}[hbt]
\centering
\includegraphics[width=\columnwidth,keepaspectratio]{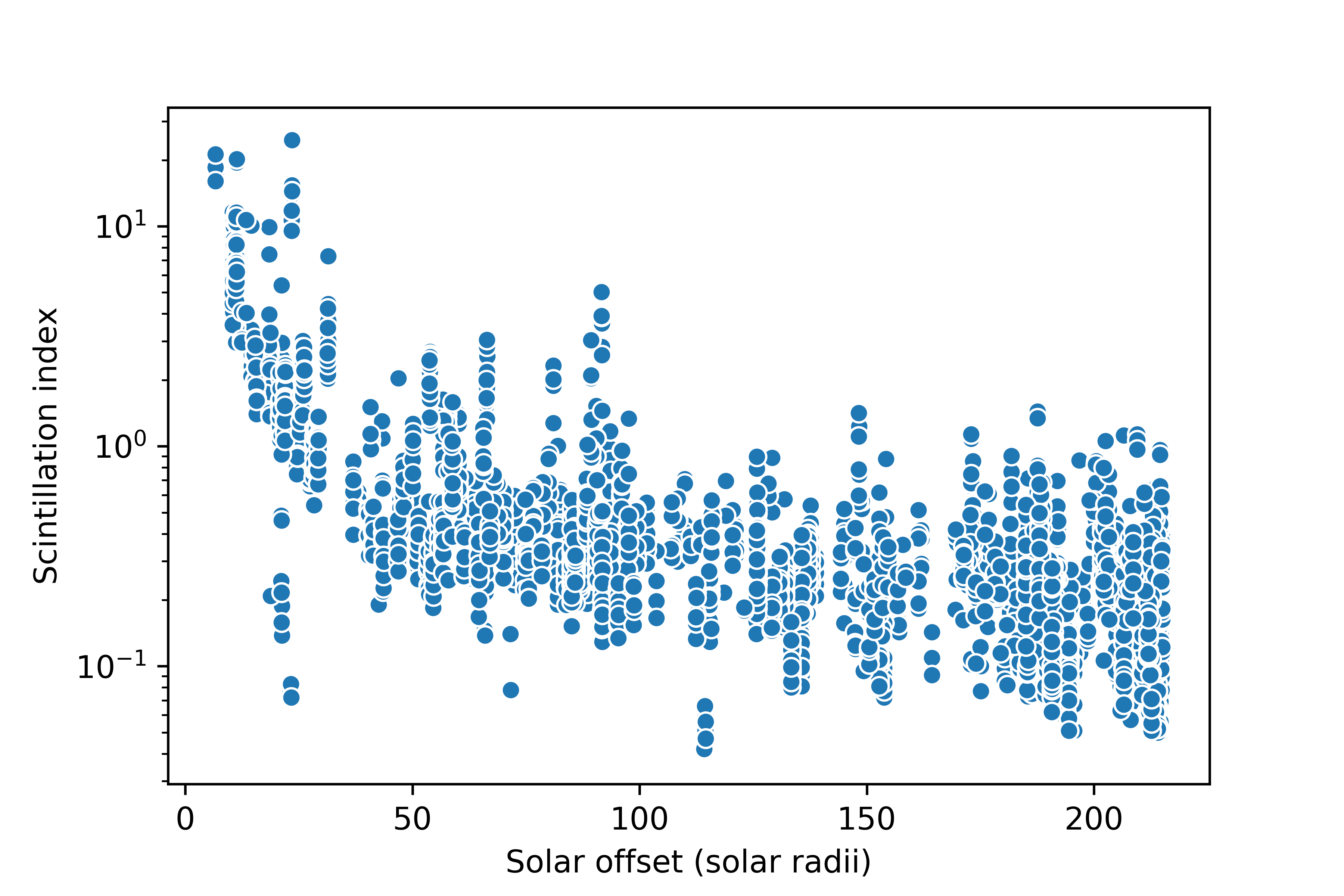}
\caption{An overview of the scintillation index variation at different solar offsets. We see that the values remain fairly low and constant at solar offsets beyond 12 solar radii. The spikes we see near 90 solar radii correspond to a coronal mass ejection (CME) event on the 6th of April 2015 \citep{molera2017analysis}. }
\label{fig:cordist}
\end{figure}

We compare the scintillation indices at various solar offsets, the distance from closest point of approach from the line of sight of observation to the Sun. This gives an insight into how the signal is affected in proximity to the Alfv\'en surface which is at nearly 12 solar radii from the Sun \citep{deforest2014inbound}. We observe the phase scintillation for different solar offsets in Figure \ref{fig:cordist}.

The intensity and phase variability are understood by calculating the power spectral density of each of the individual sessions we had. The contribution towards phase variability by each of phase scintillation and noise were calculated from within their respective bands as distinguished previously in Figure \ref{fig:spd}.

\begin{figure}
 \centering
 \includegraphics[width=\columnwidth,keepaspectratio]{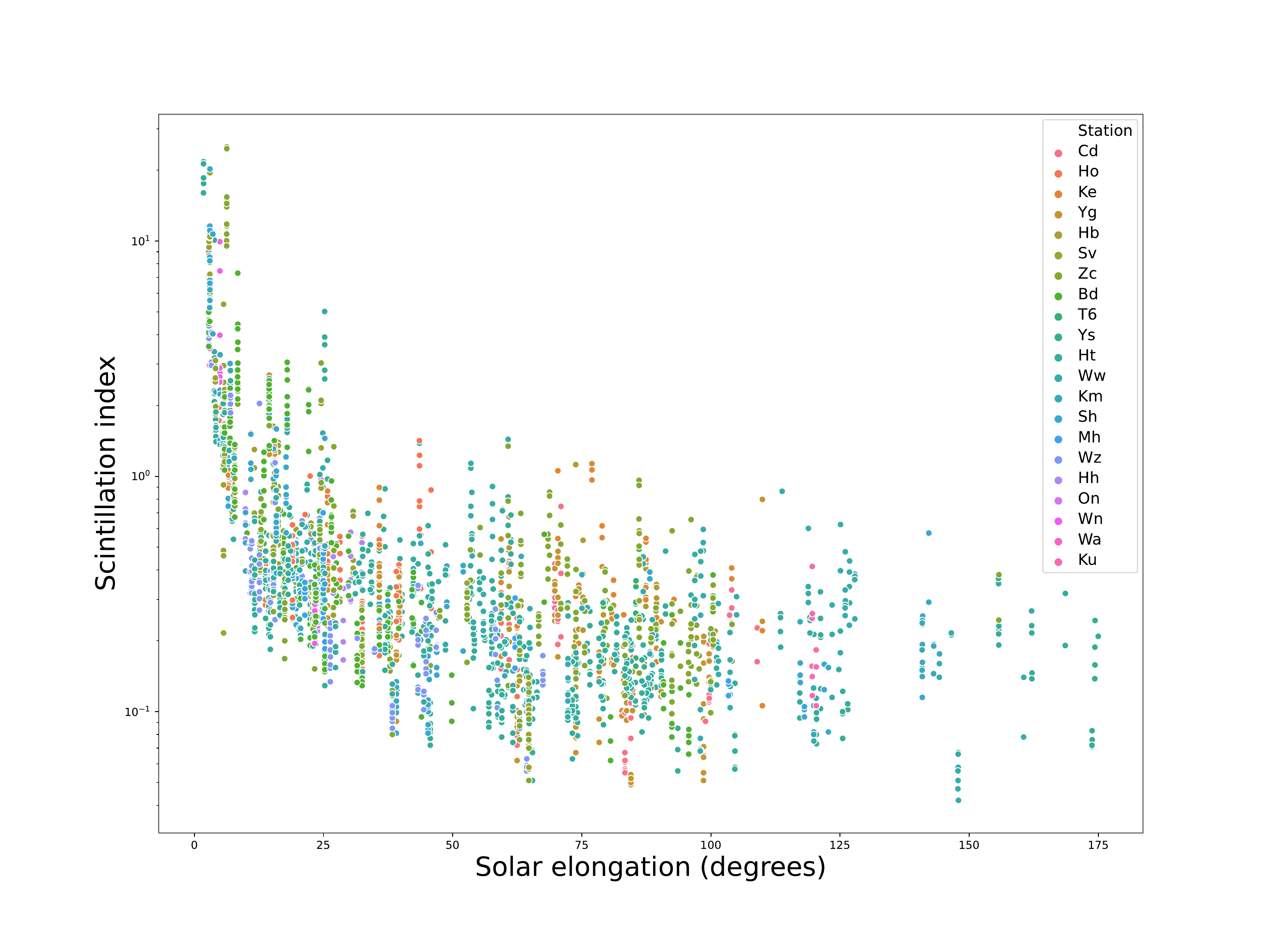}
 \caption{The scintillation index variation with solar elongation at different stations is described in the plot. The solar elongation describes the angle between the Sun, Earth, and the spacecraft; indicating that the radio signals with lower solar elongation are more closely aligned to the the Sun's emissions. We can see that there is almost a ten fold increase in the scintillation at a lower solar elongation.}
 \label{fig:sotscint}
\end{figure}

In Figure \ref{fig:sotscint}, we plot the scintillation values for each station at different solar elongation angles. The values for the scans below 10 degrees of elevation are omitted because these are saturated by the troposphere-induced scintillation. It can be seen that the lower solar offset indicates a higher perturbation in phase. Figure \ref{fig:sotscint} reiterates the idea phase fluctuations at lower solar elongation ($<$5 deg) by nearly 40 - 60 times than at higher solar elongations \citep{calves2014observations}.

\begin{figure}[hbpt]
 \centering
    \subfigure[Scintillation index]{\label{fig:scintreg}\includegraphics[width=\columnwidth,keepaspectratio]{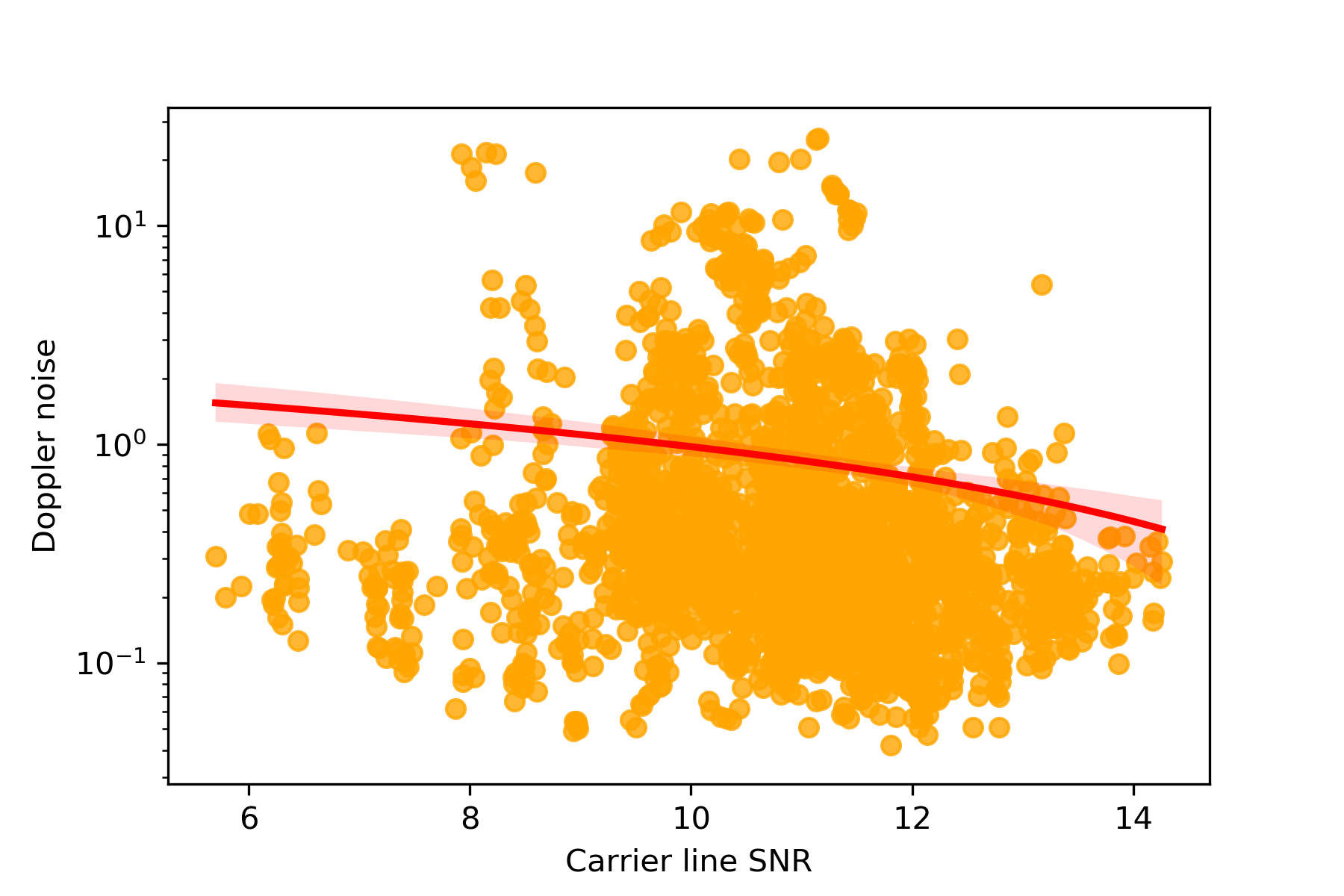}}
  \subfigure[Doppler noise]{\label{fig:dnoisereg}\includegraphics[width=\columnwidth,keepaspectratio]{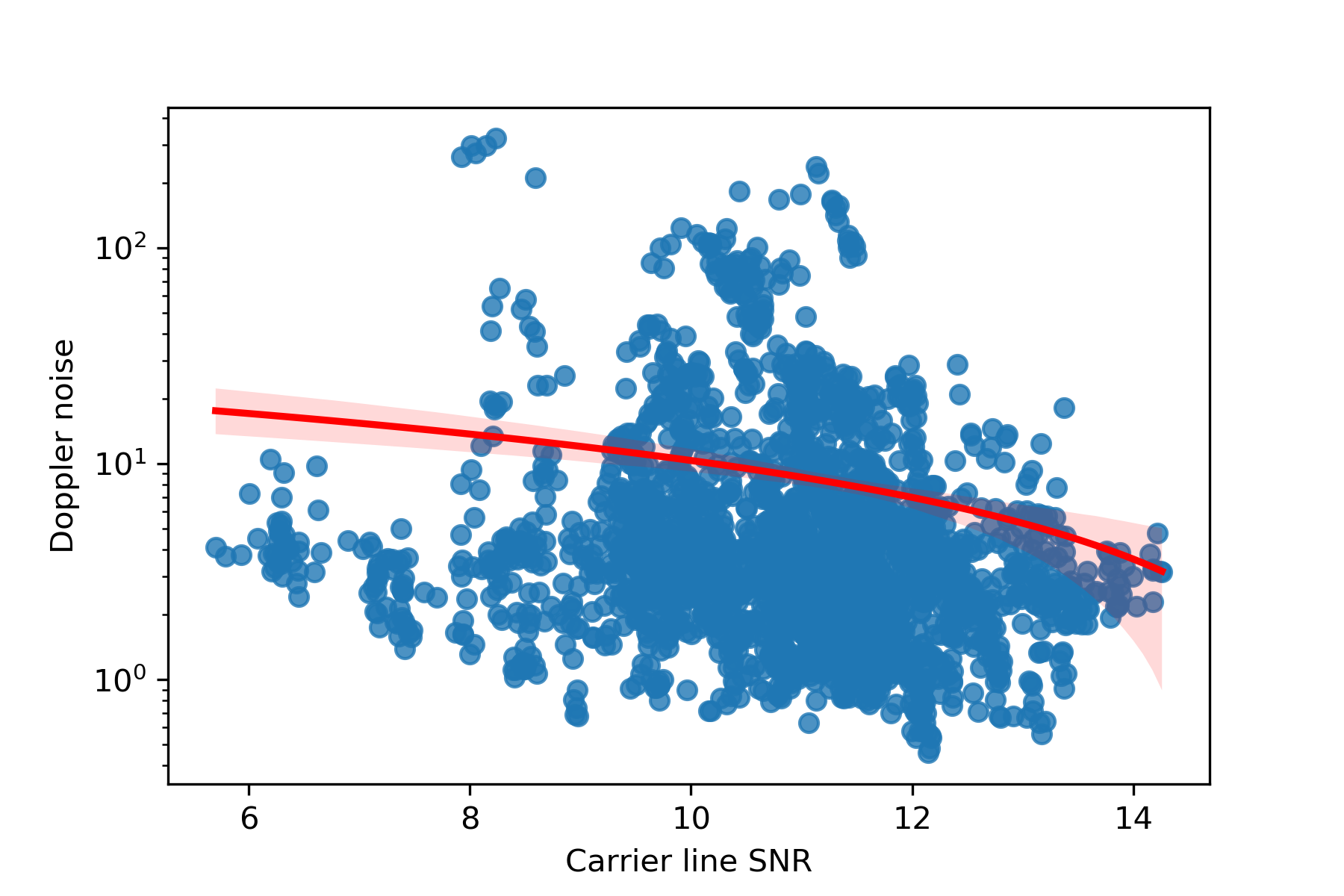}}
 \caption{The above two regression plots describe how the scintillation index (top) and the the Doppler detection noise (bottom) vary with carrier line SNR. The red line is the line with least sum of squares of errors and is the best fit for the regression.}
 \label{fig:carrsnr}
\end{figure}

The size and sensitivity of the individual telescopes do not affect the readings of the phase scintillation value. For this purpose, the carrier line signal-to-noise ratio (SNR) was compared against both the scintillation index and Doppler detection noise (2 - 10s integration time) as shown in Figure \ref{fig:carrsnr}. We performed a regression analysis to find that both the scintillation index and Doppler noise are unaffected by the carrier line SNR. The statistical test for both sets of quantities returned an r-value of -0.1 indicated they are strongly uncorrelated. The outliers we see in the Doppler Noise (above 200) are from a couple of epochs; Hartebesthoek had a session (8th June 2015) where the solar elongation was lesser than 2 degrees while Svetloe and Zelenchukskaya had a session (22nd May 2015) where the solar elongation was about 6 degrees. The scintillation index has the same sessions as the outliers (above 15) with the addition of another session from Svetloe, Zelenchukskaya and Sheshan (25th June 2015). It is possible that there was an issue with the backend baseband converters during the experiment.

The TEC of the interplanetary plasma along the line of sight from Earth to the Mars Express spacecraft is determined by integrating the electron density values obtained at points along the path of Earth and Mars for an initial value of the solar wind velocity. We aim to find the best fit of the theoretically determined model with the observations. We use a scaling factor K to relate the phase scintillation with the TEC which is an empirically determined constant lying between 2000--4000. We derive the relation from (2) to be:

\begin{ceqn}  
\begin{align}
TEC=K \cdot \sigma
\end{align}
\end{ceqn}

\begin{figure}
 \centering
 \includegraphics[width=\columnwidth,keepaspectratio]{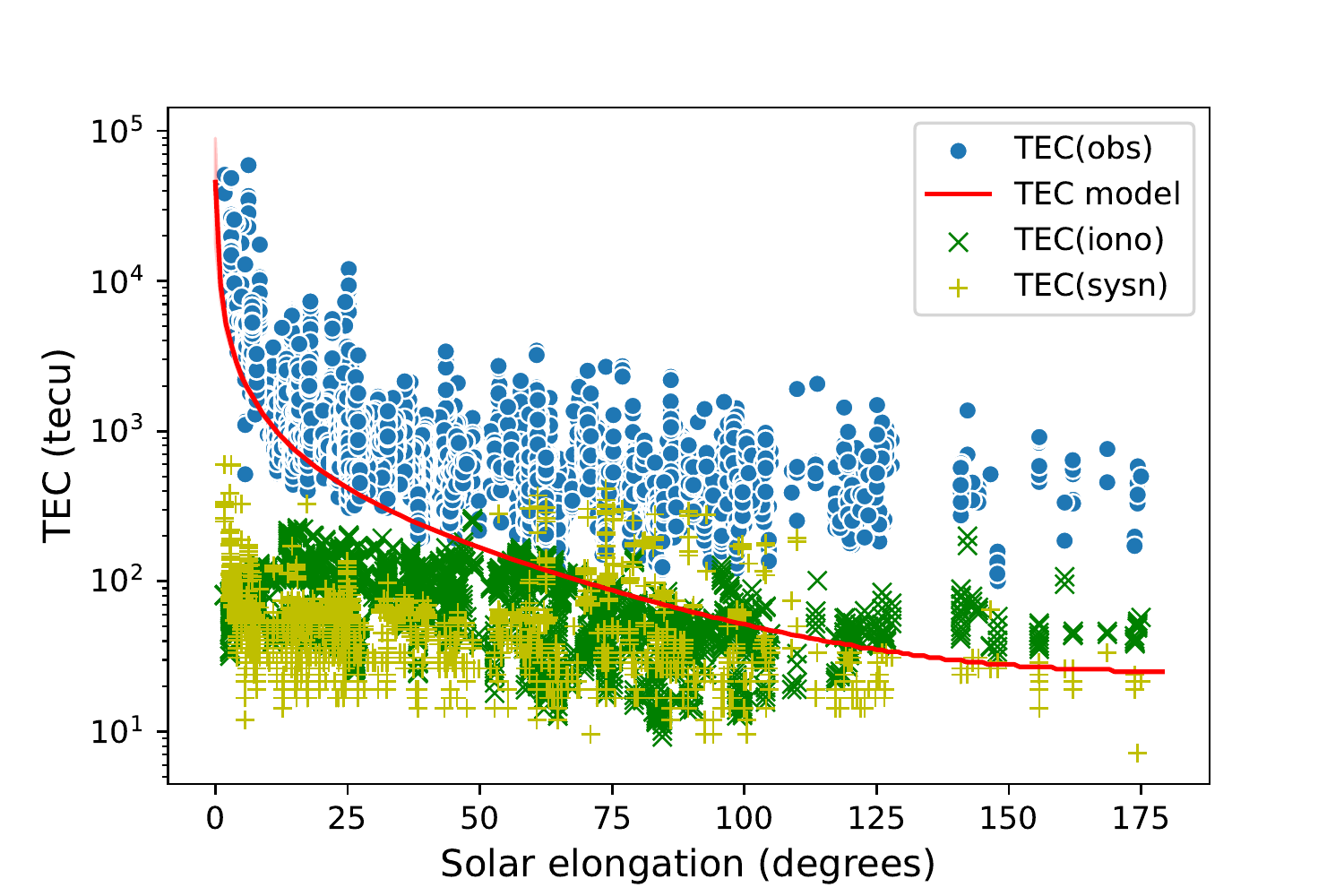}
 \caption{In this plot we see the total electron content contributions from various factors and how it compares to the theoretical fit of the model (nominal speed in this case).}
 \label{fig:tecm1}
\end{figure}

The best value for K is calculated to be 2390 from the weighted mean of the conversion factor for the best-fit scenario obtained for each point individually. The Figure \ref{fig:tecm1} depicts the TEC corresponding to various factors within our observations.

The obtained TEC value is a combination of propagative effects of the plasma, ionosphere, and mechanical and thermal noises from the instrumentation. The cumulative effect from each of these contributions can be described by the following equation

\begin{ceqn}
\begin{align}
    {Scint} = \sqrt{{TEC_{sw} \cdot k_{sw}}^2 + {TEC_{ion} \cdot k_{ion}}^2 + {PCal}^2 + {Bpr}^2 + {Airm}^2 }
\end{align}
\end{ceqn}

where $Scint$ is the measured scintillation by a single dish antenna, the $TEC_{sw} \cdot k_{sw}$ is the theoretical interplanetary two-way TEC accounting for fast, slow and mean solar speeds times a scaling factor to convert from radians per tec unit, the $TEC_{ion} \cdot k_{ion}$ is the theoretical ionospheric two-way TEC times a scaling factor to convert from radians per tec unit, $PCal$ is the instrumental phase error derived from the Phase Calibration measurements, $Bpr$ is the base phase variation due to the Allan variance of the transmission and receiving H-maser clocks, and the $Airm$ is the two-way air mass in units of $1\,kg/cm^2$

The PCal phase error was set to 0.0295 radians for those observations we did not have measurement data. The rest of the sessions the phase error was extracted directly from the measurements of these tones. We had the phase cal tones present in selected sessions to verify both independent measurements. However in standard observations they are disabled to decrease additional noise. Therefore, they are usually not estimated.

\section{Discussion}

The study involved tracking the Mars Express radio signal from 2013-2020 to study the interplanetary plasma. We measured the scintillation indices at different solar elongations from the carrier phase. The spectral index derived from the phase power spectrum returns a value of $-2.43 \pm 0.11$ which is in agreement with the turbulent media described by Kolmogorov \citep{kolmogorov1991local}.

\begin{figure}
 \centering
 \includegraphics[width=\columnwidth,keepaspectratio]{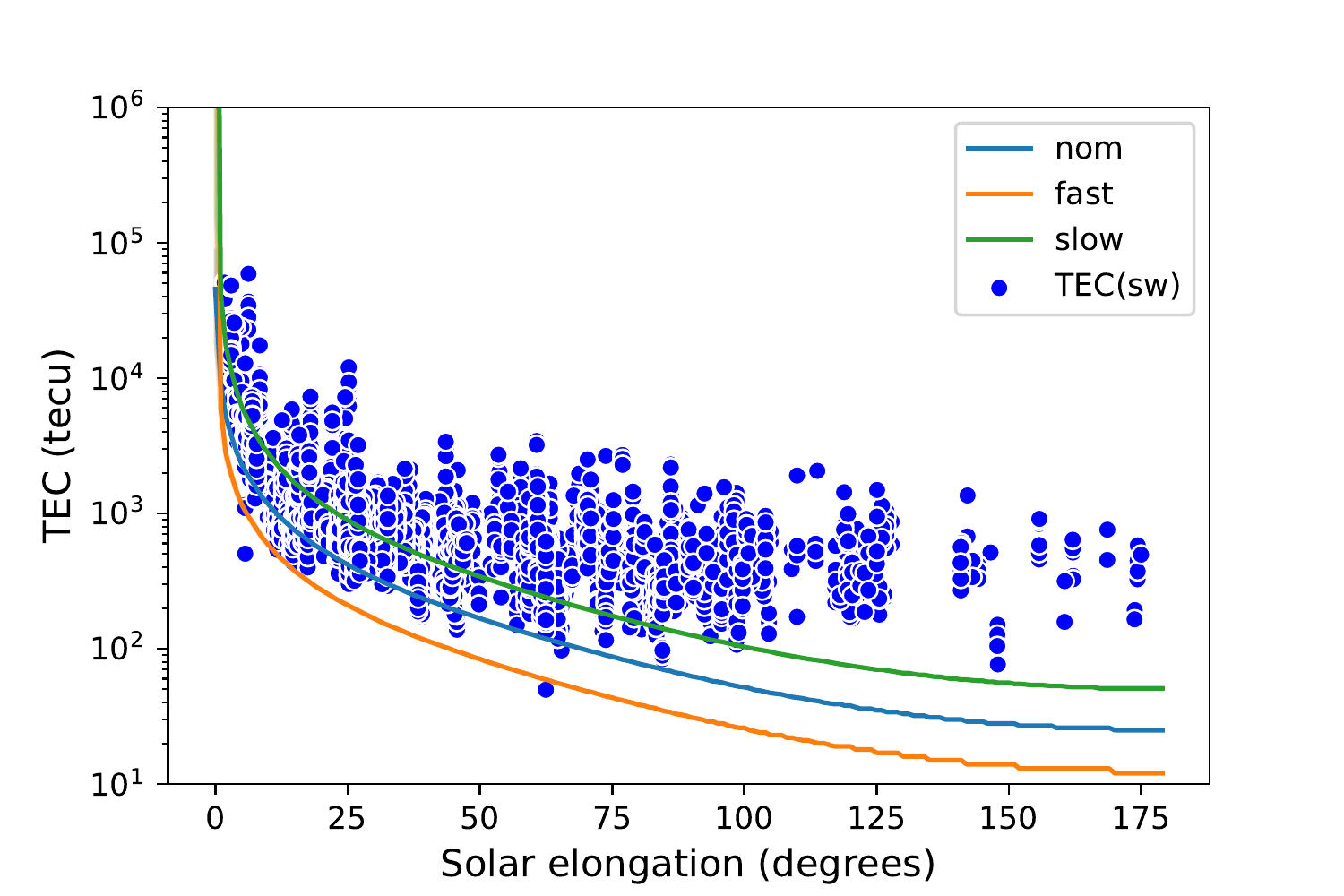}
 \caption{The blue points represent the measured values of TEC from our observations and the red line is the trend of the TEC as predicted by the theoretical model ($n_{e}=2.15\times10^6$). The prediction fails at higher elongation where the data points are significantly above the theoretical fit for nominal, slow, and fast solar wind speeds.}
 \label{fig:tecm2}
\end{figure}

We fitted the data obtained with MEX measurements to the theoretical model of TEC. We compared the results with the measurements published by VEX and improve our model. \citep{calves2014observations}. A core objective of this paper is to test our theoretical models of the total electron content by comparing to the observed total electron content. The first step was to remove the contributions from the ionosphere and the system noise because this study is concerned with the contributions due to the solar wind alone. We see an improvement of 1.8\% in the fit which shows. These newly obtained data points are overlayed with the theoretical fit as shown in Figure \ref{fig:tecm2}.

The prediction of the theoretical fit breaks down at higher solar elongations. One of the expected reasons is the speed of the solar wind which is a variant across time and direction where the ionospheric effects dominate. The second reason which could possibly explain the observed points being higher (than the prediction) at larger solar elongations is the correlation of the uplink and downlink signals. The effective variance of the two time series for the uplink and downlink signals is given as

\begin{ceqn}
\begin{align}
     {\sigma_{eff}^{2}=\sigma_{u}^{2}+\sigma_{d}^{2}+2*covariance(u,d).}
\end{align}
\end{ceqn}. 

\begin{figure}
 \centering
 \includegraphics[width=\columnwidth,keepaspectratio]{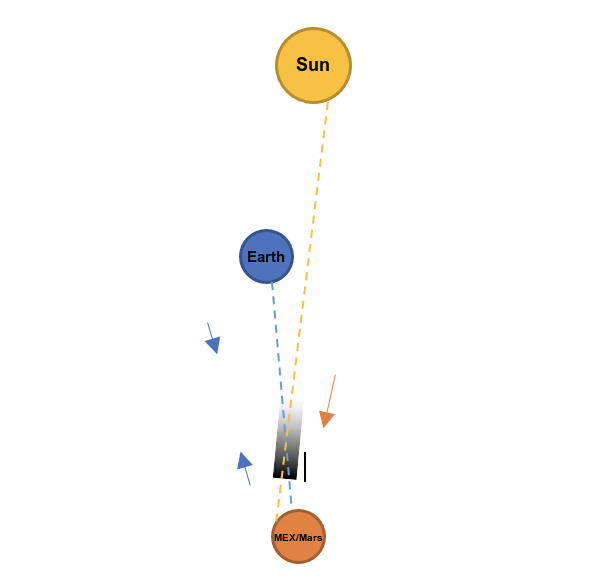}
 \caption{A two-dimensional model of transmission of the satellite signal and the solar wind direction. The yellow sphere is the Sun, the blue sphere (Earth) is the observer and the red sphere is the target (MEX). The blue arrows represent the direction of the uplink and downlink while the yellow arrow is the direction of the solar wind. The black and white slab represent the oscillating density fluctuation as a slab. The path of the solar wind would be a chain of multiple such slabs but varying in their fluctuation pattern due to solar wind's inhomogenous nature.}
 \label{fig:swin}
\end{figure}

The inhomogenous structure of plasma and the associated wind speed would mean that the uplink and downlink signal traverse through different segments of the large-scale plasma structure. Thus, the uplink and downlink contributions are uncorrelated implying $covariance(u,d)=0$. However, in cases when the solar elongation is close to 180 degrees, it is possible the solar wind ($200-800$ km/s) is moving along the direction similar to that of the signal transmission ($3\times10^5$ km/s) as depicted in Figure \ref{fig:swin}. If this is the case, the uplink and downlink signals could pass through similar plasma regions, resulting in partial correlation. This would make the covariance term non-zero leading to an increase in observed TEC value and thus explain the points being above the theoretical predictions at high solar elongations.

Having presented an improved quantitative analysis of interplanetary plasma from single dish observations, we look to further advance the technique and characterize plasma by using different line of sight observations. This would involve using multiple stations simultaneously and concurrent observations of multiple spacecraft \citep{ma2021vlbi}. The study focused on long term series analysis of phase scintillation on multiple line of sights with the same target. These phase signatures will benefit to achieve higher orbit determination accuracy on upcoming missions like JUICE. 

Another interesting domain to look into is the locational aspect of the magnetic and plasma field of the solar wind. Plasma sheets in the Sun's magnetosphere are regions of enhanced plasma with the neutral sheet; the latter are storehouses of magnetic and plasma energy released periodically \citep{mishin2021nonlinear}. The position of the spacecraft determines which regions of the magnetosphere the radio signal traverses. \cite{kim2020survey} noticed an increase in the count of ions and electrons when Juno made plasma sheet crossings. We could see minor jumps in our TEC if our radio signal made these crossings. Observing multiple targets simultaneously (MEX, BepiColombo, Tianwen \citep{ma2022detecting}, JUICE) ~\footnote{https://sci.esa.int} could provide insight into where such sheet crossings are located and is a study worth further investigating. This could consequentially help explain the increased TEC as seen in our observations compared to the theoretical model.

\begin{acknowledgement}
This study was possible thanks to the observations carried out by the different operators across the array of EVN telescopes in China, Europe, Russia, Africa, and the AUT University (for Ww and Wa). The long-term study of plasma was also consolidated by the array of the Auscope VLBI telescopes operated by the University of Tasmania. The author acknowledges the valuable input from collaborators in JIVE and Shanghai Astronomical Observatory which helped improve the quality of the work. The collection of the data for research was possible thanks to the ESA's MEX communication team. 
\end{acknowledgement}

\printendnotes

\printbibliography

\end{document}